\documentstyle[aps,prl,preprint,floats,epsfig]{revtex}

\newcommand{\goto}{\rightarrow}
\newcommand{\ra}{\rightarrow}
\newcommand{\calB}{\mbox{${\cal B}$}}
\newcommand{\calL}{\mbox{${\cal L}$}}

\newcommand{\etapr}{\mbox{$\eta^\prime$}}
\newcommand{\etaK}{\mbox{$B\ra\eta K$}}
\newcommand{\etaKst}{\mbox{$B\ra\eta K^*$}}

\newcommand{\BetaKstp}{\mbox{$\calB(B^+\ra\eta K^{*+})$}}
\newcommand{\retaKstp}{\mbox{$26.4^{+9.6}_{-8.2}\pm 3.3$}}
\newcommand{\RetaKstp}{\mbox{$(\retaKstp)\times 10^{-6}$}}
\newcommand{\etaKstz}{\mbox{$B^0\ra\eta K^{*0}$}}
\newcommand{\BetaKstz}{\mbox{$\calB(B^0\ra\eta K^{*0})$}}
\newcommand{\retaKstz}{\mbox{$13.8^{+5.5}_{-4.6}\pm 1.6$}}
\newcommand{\RetaKstz}{\mbox{$(\retaKstz)\times 10^{-6}$}}
\newcommand{\etapK}{\mbox{$B\ra\eta^\prime K$}}

\newcommand{\BetapKp}{\mbox{$\calB(B^+\ra\eta^\prime K^+)$}}
\newcommand{\retapKp}{\mbox{$80^{+10}_{-9}\pm 7$}}
\newcommand{\RetapKp}{\mbox{$(\retapKp)\times 10^{-6}$}}
\newcommand{\etapKz}{\mbox{$B^0\ra\eta^\prime K^0$}}
\newcommand{\BetapKz}{\mbox{$\calB(B^0\ra\eta^\prime K^0)$}}
\newcommand{\retapKz}{\mbox{$89^{+18}_{-16}\pm 9$}}
\newcommand{\RetapKz}{\mbox{$(\retapKz)\times 10^{-6}$}}

\newcommand{\DE}{\mbox{$\Delta E$}}
\newcommand{\mb}{\mbox{$M$}}
\newcommand{\xf}{\mbox{${\cal F}$}}
\newcommand{\hel}{\mbox{${\cal H}$}}
\newcommand{\piz}{\mbox{$\pi^0$}}
\def\sgline{\noalign{\vskip 0.15truecm\hrule\vskip 0.15truecm}}

\newcommand{\etaprkp}{\mbox{$\etapr K^+$}}
\newcommand{\etaprkpd}{\mbox{$\etapr_{\eta\pi\pi}K^+$}}
\newcommand{\etaprkprg}{\mbox{$\etapr_{\rho\gamma~}K^+$}}

\newcommand{\etaprkz}{\mbox{$\etapr K^0$}}
\newcommand{\etaprkzd}{\mbox{$\etapr_{\eta\pi\pi} K^0$}}
\newcommand{\etaprkzrg}{\mbox{$\etapr_{\rho\gamma~} K^0$}}
\newcommand{\etaprpi}{\mbox{$\etapr\pi^+$}}
\newcommand{\etaprpid}{\mbox{$\etapr_{\eta\pi\pi}\pi^+$}}
\newcommand{\etaprpirg}{\mbox{$\etapr_{\rho\gamma~}\pi^+$}}

\newcommand{\etaprpiz}{\mbox{$\etapr\piz$}}
\newcommand{\etaprpizepp}{\mbox{$\etapr_{\eta\pi\pi}\piz$}}
\newcommand{\etaprpizrg}{\mbox{$\etapr_{\rho\gamma~}\piz$}}
\newcommand{\etaprkstz}{\mbox{$\etapr K^{*0}$}}
\newcommand{\etaprkstzd}{\mbox{$\etapr_{\eta\pi\pi} K^{*0}$}}
\newcommand{\etaprkstp}{\mbox{$\etapr K^{*+}$}}
\newcommand{\etaprkstpd}{\mbox{$\etapr_{\eta\pi\pi} K^{*+}_{K^+\piz}$}}
\newcommand{\etaprkstpkz}{\mbox{$\etapr_{\eta\pi\pi} K^{*+}_{K^0\pi^+}$}}
\newcommand{\etaprkstzrg}{\mbox{$\etapr_{\rho\gamma~} K^{*0}$}}
\newcommand{\etaprkstprg}{\mbox{$\etapr_{\rho\gamma~} K^{*+}_{K^+\piz}$}}
\newcommand{\etaprkstpkzrg}{\mbox{$\etapr_{\rho\gamma~} K^{*+}_{K^0\pi^+}$}}
\newcommand{\etaprrhozrg}{\mbox{$\etapr_{\rho\gamma~}\rho^0$}}
\newcommand{\etaprrhoprg}{\mbox{$\etapr_{\rho\gamma~}\rho^+$}}
\newcommand{\etaprrhoz}{\mbox{$\etapr\rho^0$}}
\newcommand{\etaprrhozd}{\mbox{$\etapr_{\eta\pi\pi}\rho^0$}}
\newcommand{\etaprrhop}{\mbox{$\etapr\rho^+$}}
\newcommand{\etaprrhopd}{\mbox{$\etapr_{\eta\pi\pi}\rho^+$}}
\newcommand{\etak}{\mbox{$\eta K^+$}}
\newcommand{\etakgg}{\mbox{$\eta_{\gaga} K^+$}}
\newcommand{\etakthrp}{\mbox{$\eta_{3\pi} K^+$}}
\newcommand{\etapi}{\mbox{$\eta\pi^+$}}
\newcommand{\etapigg}{\mbox{$\eta_{\gaga}\pi^+$}}
\newcommand{\etapithrp}{\mbox{$\eta_{3\pi}\pi^+$}}
\newcommand{\etapiz}{\mbox{$\eta\piz$}}
\newcommand{\etapizgg}{\mbox{$\eta_{\gaga}\piz$}}
\newcommand{\etapizthrp}{\mbox{$\eta_{3\pi}\piz$}}
\newcommand{\etakz}{\mbox{$\eta K^0$}}
\newcommand{\etakzgg}{\mbox{$\eta_{\gaga} K^0$}}
\newcommand{\etakzthrp}{\mbox{$\eta_{3\pi} K^0$}}

\newcommand{\etakstz}{\mbox{$\eta K^{*0}$}}
\newcommand{\etakstzgg}{\mbox{$\eta_{\gaga} K^{*0}$}}
\newcommand{\etakstzthrp}{\mbox{$\eta_{3\pi} K^{*0}$}}
\newcommand{\etakstp}{\mbox{$\eta K^{*+}$}}
\newcommand{\etakstpgg}{\mbox{$\eta_{\gaga} K^{*+}_{K^+\piz}$}}
\newcommand{\etakstpthrp}{\mbox{$\eta_{3\pi} K^{*+}_{K^+\piz}$}}
\newcommand{\etakstpggkz}{\mbox{$\eta_{\gaga} K^{*+}_{K^0\pi^+}$}}
\newcommand{\etakstpthrpkz}{\mbox{$\eta_{3\pi} K^{*+}_{K^0\pi^+}$}}
\newcommand{\etarhoz}{\mbox{$\eta \rho^0$}}
\newcommand{\etarhozgg}{\mbox{$\eta_{\gaga} \rho^0$}}
\newcommand{\etarhozthrp}{\mbox{$\eta_{3\pi} \rho^0$}}
\newcommand{\etarhop}{\mbox{$\eta \rho^+$}}
\newcommand{\etarhopgg}{\mbox{$\eta_{\gaga} \rho^+$}}
\newcommand{\etarhopthrp}{\mbox{$\eta_{3\pi} \rho^+$}}
\newcommand{\gaga}{{\gamma\gamma}}

\begin{document}

\preprint{\tighten\vbox{\hbox{\hfil CLNS 99/1649}
                        \hbox{\hfil CLEO 99-16}
}}

\title{Two-body \boldmath{$B$} Meson Decays to \boldmath{$\eta$} and
\boldmath{$\eta^\prime$}: Observation of \boldmath{$B\ra\eta K^*$}
}  

\author{CLEO Collaboration}
\date{\today}

\maketitle
\tighten

\begin{abstract} 

In a sample of 19 million produced $B$ mesons, we have observed the decays
\etaKst\ and improved our previous measurements of \etapK.  The 
branching fractions we measure for these decay modes are $\BetaKstp =
\RetaKstp$, $\BetaKstz = \RetaKstz$, $\BetapKp = \RetapKp$ and\linebreak
$\BetapKz = \RetapKz$.  We have searched with comparable sensitivity for
related decays and report upper limits for these branching fractions.

\end{abstract}

\newpage

{
\renewcommand{\thefootnote}{\fnsymbol{footnote}}

% Insert author and address list here

\begin{center}
S.~J.~Richichi,$^{1}$ H.~Severini,$^{1}$ P.~Skubic,$^{1}$
A.~Undrus,$^{1}$
S.~Chen,$^{2}$ J.~Fast,$^{2}$ J.~W.~Hinson,$^{2}$ J.~Lee,$^{2}$
N.~Menon,$^{2}$ D.~H.~Miller,$^{2}$
% ...
\end{center}
 
\small
\begin{center}
$^{1}${University of Oklahoma, Norman, Oklahoma 73019}\\
$^{2}${Purdue University, West Lafayette, Indiana 47907}\\
% ...
\end{center}

\setcounter{footnote}{0}
}
\newpage

% Insert body of the text here.

\begin{figure}[htbp]
\setlength{\epsfxsize}{1.0\linewidth}\leavevmode\epsfbox{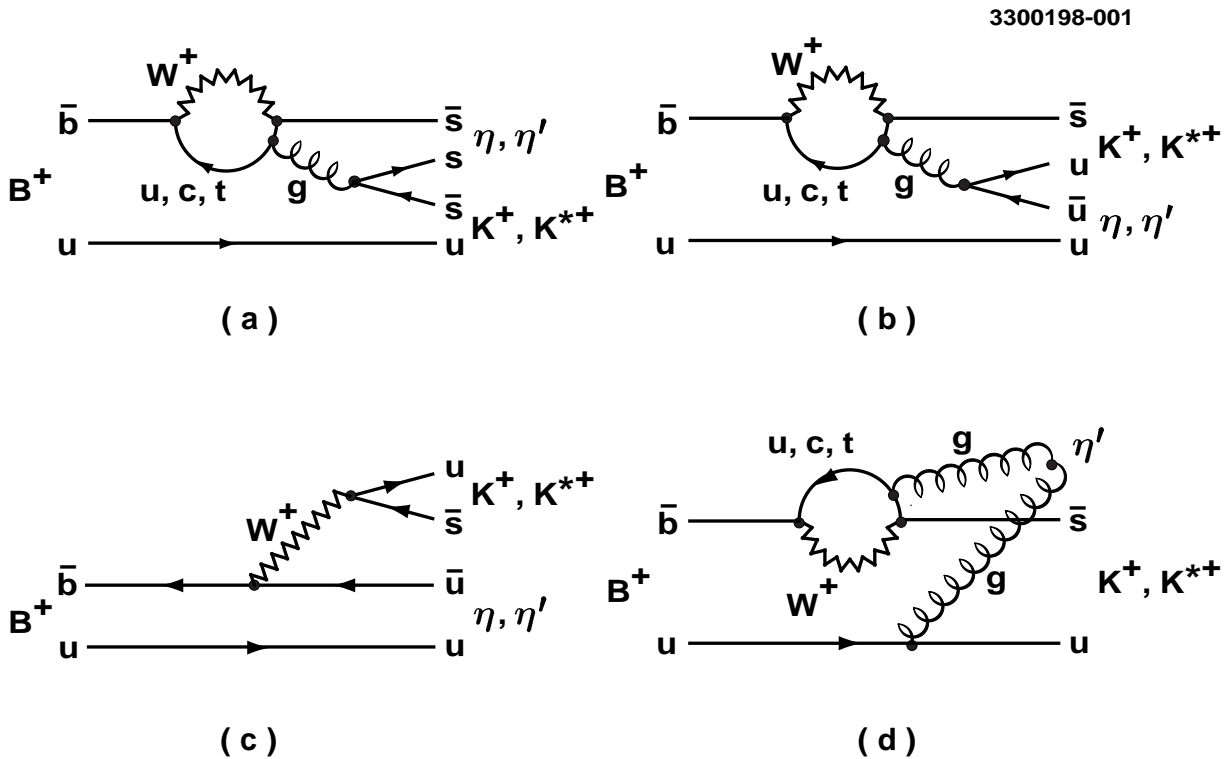}
\caption{\label{fig:diagrams}
Feynman diagrams describing the representative decays
$B^+\ra\eta^{(\prime)}K^{(*)+}$:  (a, b) internal penguins; (c)
external spectator; (d) flavor-singlet penguin.}  
\end{figure}

There has been considerable recent interest in charmless hadronic $B$
decays, partly because of the observation of several of these
decays \cite{Bkpi,Betapr,Bomegaphi}, and partly because of their
anticipated importance in understanding the phenomenon
of $CP$ violation.  These decays are expected to proceed primarily through
$b\ra s$ loop (``penguin") diagrams and $b\ra u$ spectator diagrams.  In
Fig. \ref{fig:diagrams} we show four such diagrams which may be expected
to contribute to the decays involving isoscalar mesons which are the
subject of this Letter.  
An earlier search \cite{Betapr} found a large rate for the decay \etapK, and set
upper limits on other decays to two-body final states containing $\eta$
or $\etapr$ mesons.  
Recent effective Hamiltonian predictions 
\cite{AKL,CCTY} of the decay branching fractions of the \etapK\ decay are still
somewhat smaller than the measurement \cite{Betapr}.

We present results of improved experimental searches
for $B$ meson decays to two-body final states containing $\eta$ or
$\eta^\prime$ mesons
with the first observation of the decay \etaKst.
These results are based on data collected with the CLEO II
detector~\cite{CLEOdet} at the Cornell Electron Storage Ring (CESR). The data
sample corresponds to an integrated luminosity of 9.13~fb$^{-1}$ for the
reaction $e^+ e^- \rightarrow \Upsilon(4S) \rightarrow B\overline{B}$, which
in turn corresponds to $9.66\times 10^6$ $B\overline{B}$ pairs
\cite{chgneutfrac}.  To study
background from continuum processes, we also collected 4.35~fb$^{-1}$ of data
at a center-of-mass energy below the threshold for $B\overline{B}$ production.
These constitute the complete data sample from the
CLEO II and CLEO II.V experiments, and the measurements we report here
supersede our earlier results \cite{Betapr}\ from a subset of these data.

The CLEO II detector emphasizes precision charged
particle tracking, with specific ionization ($dE/dx$) measurement, and
high resolution electromagnetic calorimetry based on CsI(Tl).
Scintillators between the tracking chambers and calorimeter provide 
time-of-flight (TOF) information which we use in conjunction with $dE/dx$ for
particle identification (PID).  The CLEO II.V 
configuration \cite{CLEO2.5} differs in two respects: the replacement of an
inner straw-tube drift chamber with a three-layer, double-sided-silicon 
vertex detector; and replacement of the 50:50 argon-ethane gas
in the main drift chamber with a 60:40 helium-propane mixture.  

We reconstruct charged pions and kaons, photons, 
and $\pi^+\pi^-$ pairs that intersect
at a vertex displaced from the
collision point (``vees'' from $K^0_S\ra\pi^+\pi^-$).  Candidate $B$
decay tracks must meet specifications on the number of drift chamber
measurements, goodness of fit, and consistency with an origin at the
primary or particular secondary vertex.  
Candidate photons (from $\pi^0$, $\eta$, and $\eta^\prime$ decays) must be
isolated calorimeter clusters with a photon-like spatial distribution
and energy deposition exceeding 30 MeV.  In order to reject soft photon
backgrounds, we require $\eta\to\gamma\gamma$ candidates to satisfy 
$\left|\cos\theta^*\right|<0.97$, where
$\theta^*$ is the center-of-mass decay angle relative to its
flight direction.  This cut is tightened to 0.90
for $\eta K^*/\rho$ channels to veto $B\ra K^*\gamma$
background.  We reject charged tracks and photon pairs having
momentum less than 100 MeV/c.  The photon from candidate
$\etapr\ra\rho\gamma$ decays is required to have an energy
greater than 200 MeV, though this requirement is relaxed to 100 MeV for
channels with relatively low background.

We fit photon pairs and vees kinematically to the appropriate combined
mass hypothesis to obtain meson momenta. 
The reconstructed mass resolutions prior to the constraint are about 5--10
MeV (momentum dependent)
for $\pi^0\ra\gamma\gamma$, 12 MeV for $\eta\ra\gamma\gamma$,
and 3 MeV for $K^0_S\ra\pi^+\pi^-$.  We determine the expected signal
distributions for these and other quantities needed in the analysis from
a detailed $\scriptstyle\rm GEANT$ based simulation of the CLEO detector
\cite{GEANT}\ and studies of data for a variety of benchmark processes.
In particular, we have determined the momentum and $dE/dx$
resolutions in studies of $D^0\ra K^-\pi^+$ data events 
for track momenta greater than 2.0 GeV/c.

 The primary means of identification of $B$ meson candidates is through their
 measured mass and energy. The quantity $\Delta E$ is defined as $\Delta E
 \equiv E_1 + E_2 - E_b$, where $E_1$ and $E_2$ are the energies of the two
 $B$ daughters (typically $\sim$2.6 GeV) and $E_b$ is the beam energy 
 (5.29 GeV). The
 beam-constrained mass of the candidate is defined as $M \equiv \sqrt{E_b^2 -
 |{\bf p}|^2}$, where $\bf p$ is the measured momentum of the candidate
 (typically $|{\bf p}|$ $\sim$ 325 MeV/c).  We use
 the beam energy instead of the measured energy of the $B$ candidate to
 improve the mass resolution by about one order of magnitude.

For vector-pseudoscalar decays of the $B$ and the $\rho\gamma$ decay of the
$\eta^\prime$, we gain further discrimination from the
helicity variable \hel, the cosine of the vector meson's rest frame
two-body decay angle with respect to its flight direction, which
reflects the spin alignment in the decay. 
The decay rate is proportional to $\hel^2$ when the vector meson decays
into two spinless daughters, and to $1-\hel^2$ for $\etapr\ra\rho\gamma$.
For modes in which one daughter is a single
charged track,
$dE/dx$ measurements provide statistical discrimination between kaons and pions.
With $S_K$ and $S_\pi$ defined as the
deviations from nominal energy loss for the indicated particle
hypotheses measured in standard deviations, the separation $S_K-S_\pi$
is about 1.7 (2.0) at 2.6 GeV/c for the CLEO II (II.V) samples. 
										
 The large background from continuum quark--antiquark ($q\bar q$) production
 can be reduced with event shape cuts. Because $B$ mesons are
 produced almost at rest, the decay products of the $B\bar B$ pair tend to be
 isotropically distributed, while particles from $q\bar q$ production have a
 more jet-like distribution. The angle $\theta_{T}$ between the thrust axis of
 the charged particles and photons forming the candidate $B$ and the thrust
 axis of the remainder of the event is required to satisfy $|\cos{\theta_{T}}|
 < 0.9$. Continuum background is strongly peaked near 1 and signal is
 approximately flat for $|\cos{\theta_{T}}|$. We also form a multivariate 
 discriminant (${\cal F}$)~\cite{bigrare} from the momentum scalar sum of
 charged particles and photons in nine cones of increasing polar angle around
 the thrust axis of the candidate, the angle of the thrust axis of the
 candidate, and the direction of $\bf p$ with respect to the beam axis.  
 We have checked the backgrounds from the dominant $B$ decay modes 
 ($b\to c$) by simulation, finding their
 contributions to the modes in this study to be generally quite small.
 Where appropriate we include this component in the fits described below.  

The selection criteria for mass, energy, and event shape variables are
chosen to include sidebands about the expected signal peaks. 
To extract event yields we perform unbinned extended maximum likelihood
fits to the data \cite{Betapr}.
Observables for each event include \mb,
\DE, \xf, and (where applicable) resonance masses and \hel.  The number of 
events included in the fits ranges from $\sim$100 to 20,000.

For $B^+$ decays \cite{chgconj}\ that
have a primary daughter charged hadron (generically
$h^+$) that can be either $\pi^+$ or $K^+$, we fit both modes simultaneously,
with the likelihood \calL\ expanded so that the signal and background yields 
of both $\pi^+$ and $K^+$ are fit variables.  The modes with a
secondary vector decay involving $h^+$ ($K^*\ra K^+\pi$ and $\rho\ra\pi^+\pi$)
also require special treatment.  For these modes the momentum
spectrum of $h^+$ is bimodal because of the forward-backward peaked
\hel\ distribution.  We select independent
$K^*$ and $\rho$ samples to fit according to the sign of $\hel$.  
Events with $\hel<0$ in our sign convention
have low momentum $h^+$ and are unambiguously separated
by kinematics combined with PID information from $dE/dx$ measurements.  
For the events with $\hel>0$ the separation is much smaller, so we fit both
$K^*$ and $\rho$ yields simultaneously, using the $K^*$ hypothesis for \hel.
In all cases involving two $h^+$ hypotheses, we
include the normalized $dE/dx$ observables $S_\pi$ and $S_K$ in the fit.
For $K^{*0}$, we distinguish $K^+\pi^-$ from $K^-\pi^+$ candidates
using $dE/dx$ and TOF information.  The kinematics and the definition 
of \hel\ for these neutral decays causes $\sim$85\% of all
$\rho^0\to\pi^+\pi^-$ signal candidates to be assigned to the $\hel>0$ sample.
All possible combinations are included except $K^{*0}\to K^0\pi^0$
(efficiency too small) and $\hel>0$ with low momentum $\pi^0$ for 
the $\eta^\prime\ra\rho\gamma$ channel (background too large).

The probability distribution functions (PDF) are constructed as products 
of functions of the observables.  
For signal the dependences on masses and energies 
are represented by Gaussian, double Gaussian, or Breit-Wigner
functions, whose parameters are fixed
in the fit.  The background PDF contains signal-like peaking 
components in its resonance mass projections, to account for real
resonances in the background, added to smooth components for
combinatoric continuum.  The smooth components are low-order polynomials,
except that for \mb\ we use an empirical shape
\cite{argus}\ that accounts for the phase space limit at $M=E_b$.  
The signal and background dependences of \xf, $S_K$, and
$S_\pi$ are bifurcated Gaussian functions.  We obtain the signal parameters
from separate fits to simulated signal, and background parameters
from fits to the below-threshold data sample.
If the simulation estimate of background from $\Upsilon(4S)$
production is non-negligible, we add a term with a free fit variable to
account for this as well.

Intermediate results for all of the $B$ decay chains appear
in Table \ref{individtab}.  
Where relevant, the two \hel\ hemispheres have been combined.
We combine the samples from multiple secondary decay channels 
by adding the $-2\ln \calL$ 
functions of branching fraction and extracting a value with errors
or 90\% confidence level (CL) upper limit from the combined distribution.  
The limit is the value of
\calB\ below which lies 90\% of the integral of \calL.
In Table \ref{combtab}\ we summarize the final results for these measurements
with theoretical estimates \cite{theory}.
The first error is statistical and the second systematic.  
The latter include systematic contributions from
uncertainties in the PDFs \cite{Betapr},
reconstruction efficiencies and selection requirements ($\sim$10--15\%).
We quote 
limits computed with efficiencies reduced by one standard deviation.

\begin{table}[t!]
\tighten
\vbox{
\caption{Intermediate results for final states listed in the first column,
with the subscripts denoting secondary
decays, including $\eta^\prime\ra \eta\pi^+\pi^-$ ($\eta\pi\pi$) with
$\eta\ra\gamma\gamma$ ($\gamma\gamma$), 
$\eta^\prime\ra \rho\gamma$ ($\rho\gamma$), 
and $\eta\ra\pi^+\pi^-\pi^0$ ($3\pi$).  
The remaining columns give event yield from the fit, 
reconstruction efficiency $\epsilon$, total efficiency with secondary
branching fractions ${\cal B}_s$, and the resulting $B$
decay branching fraction ${\cal B}$, with statistical error only.
}
\begin{center}
\begin{tabular}{lcrrccc}
Final state & Fit events & $\epsilon$(\%) & $\epsilon\calB_s$(\%) &
              \calB($10^{-6})$ \cr
\sgline
\etaprkpd     & $39.6^{+7.0}_{-6.4}$ & 27 & 4.7 & 
                $88^{+16}_{-14}$ \cr
\etaprkprg  & $61^{+11}_{-10}$ & 29 & 8.7 & 
                $72^{+13}_{-12}$ \cr
\etaprkzd      & $9.2^{+3.6}_{-2.9}$ & 24 & 1.4 &
                $67^{+26}_{-21}$ \cr
\etaprkzrg    & $29.6^{+7.0}_{-6.2}$ & 28 & 2.9 &
               $105^{+25}_{-22}$ \cr
\etaprpid      & 
$0.0^{+2.2}_{-0.0}$ & 28 & 4.7 & $0.0^{+4.9}_{-0.0}$ \cr
\etaprpirg     & 
$4.4^{+7.2}_{-4.4}$ & 30 & 9.0 & $5.1^{+8.3}_{-5.1}$ \cr
\etaprpizepp      & 
$0.0^{+0.6}_{-0.0}$ & 23 & 4.0 & $0.0^{+1.5}_{-0.0}$ \cr
\etaprpizrg       & 
$0.8^{+4.1}_{-0.8}$ & 27 & 8.3 & $1.0^{+5.2}_{-1.0}$ \cr
$\etaprkstpd$ & 
$0.0^{+2.3}_{-0.0}$ & 14 & 0.8 & $0^{+30}_{-0}$ \cr
$\etaprkstprg$ &
$0.1^{+3.3}_{-0.1}$ & ~9 & 0.9 & $1^{+39}_{-1}$ \cr
\etaprkstpkz &
$0.0^{+0.6}_{-0.0}$ & 16 & 0.6 & $0^{+10}_{-0}$ \cr
\etaprkstpkzrg &
$3.2^{+2.9}_{-1.9}$ & 19 & 1.3 & $25^{+23}_{-15}$ \cr
$\etaprkstzd$ &
$2.4^{+2.7}_{-1.6}$ & 20 & 2.3 & $11^{+12}_{-7}$ \cr
$\etaprkstzrg$ &
$0.0^{+3.4}_{-0.0}$ & 21 & 4.1 & $0^{+8.7}_{-0}$ \cr
$\etaprrhopd$ &
$2.6^{+2.8}_{-1.5}$ & 15 & 2.5 & $11^{+12}_{-6}$  \cr
$\etaprrhoprg$ &
$3.2^{+6.7}_{-3.2}$ & ~9 & 2.7 & $12^{+26}_{-12} $ \cr
\etaprrhozd& 
$0.0^{+0.9}_{-0.0}$ & 17 &  2.9 & $0.0^{+3.2}_{-0.0} $ \cr
\etaprrhozrg&
$2.2^{+4.3}_{-2.2}$ & 17 &  5.1 & $4.4^{+8.7}_{-4.4} $ \cr
\etakgg   &
$5.9^{+6.0}_{-4.6}$ & 45 & 17.5 & $3.5^{+3.5}_{-2.7}$  \cr
\etakthrp &
$0.0^{+2.0}_{-0.0}$ & 29 &  6.6 & $0.0^{+3.1}_{-0.0}$  \cr
\etakzgg  &
$0.0^{+2.6}_{-0.0}$ & 38 &  5.1 & $0.0^{+5.2}_{-0.0}$  \cr
\etakzthrp     &
$0.0^{+0.9}_{-0.0}$ & 25 &  1.9 & $0.0^{+5.0}_{-0.0}$  \cr
\etapigg  &
$5.7^{+5.7}_{-4.6}$ & 46 & 18.2 & $3.2^{+3.3}_{-2.6}$  \cr
\etapithrp &
 $0.0^{+1.1}_{-0.0}$& 30 &  6.8 & $0.0^{+1.7}_{-0.0}$  \cr
\etapizgg &
 $0.0^{+1.0}_{-0.0}$& 35 & 13.7 & $0.0^{+0.8}_{-0.0}$  \cr
\etapizthrp    &
 $0.0^{+1.4}_{-0.0}$& 20 &  4.6 & $0.0^{+3.1}_{-0.0}$  \cr
$\etakstpgg$ & 
$9.3^{+5.2}_{-3.5}$ & 22 & 2.8 & $34^{+19}_{-13}$ \cr
$\etakstpthrp$ &
$3.6^{+3.1}_{-2.3}$ & 15 & 1.1 & $32^{+28}_{-20}$ \cr
\etakstpggkz &
$3.3^{+3.0}_{-2.1}$ & 25 & 2.2 & $16^{+14}_{-10}$ \cr
\etakstpthrpkz &
$3.0^{+2.7}_{-1.9}$ & 17 & 0.9 & $34^{+30}_{-21}$ \cr
$\etakstzgg$ &
$7.8^{+4.7}_{-3.1}$ & 32 & 8.3 & $9.7^{+5.8}_{-3.9}$ \cr
$\etakstzthrp$ &
$8.0^{+4.4}_{-3.5}$ & 21 & 3.3 & $25^{+14}_{-11}$ \cr
$\etarhopgg$ &
$0.0^{+2.5}_{-0.0}$ & 22 & 8.5 & $0.0^{+3.1}_{-0.0}$ \cr
$\etarhopthrp$ &
$5.0^{+4.6}_{-3.0}$ & 15 & 3.4 & $15^{+14}_{-9}$ \cr
\etarhozgg & 
$2.0^{+3.2}_{-2.4}$& 26 & 10.3 & $~2.0^{+3.3}_{-2.0}$ \cr
\etarhozthrp &
$2.3^{+4.3}_{-2.3}$& 18 &  4.2 & $~6^{+11}_{-6}$ 
\end{tabular}
\end{center}
\label{individtab}
}
\end{table}

\begin{table}[htbp]
\tighten
\caption{Combined branching fractions ($\calB_{\rm fit}$), significance and 
final result ($\calB$).  The statistical and systematic errors are given for
$\calB_{\rm fit}$ except where the result is not statistically
significant, in which case they are combined and the final result is 
quoted as a 90\%
confidence level upper limit.  We quote estimates from various theoretical
sources \protect\cite{theory} for comparison. }
\begin{center}
\begin{tabular}{lccccl}
Decay mode & $\calB_{\rm fit}(10^{-6})$ & Signif. & $\calB(10^{-6})$ 
                                        & Theory \cr
 & & ($\sigma$) & & \calB($10^{-6}$) \cr
\sgline
$B^+\goto\etaprkp$    & \retapKp & 16.8 &   see $\calB_{\rm fit}$   &
                         7--65 \cr
$B^0\goto\etaprkz$    & \retapKz & 11.7 &   see $\calB_{\rm fit}$   &
                         9--59 \cr
$B^+\goto\etaprpi$    & $1.0^{+5.8}_{-1.0}$   & 0.0 & $<12$ &
                         1--23 \cr
$B^0\goto\etaprpiz$   & $0.0^{+1.8}_{-0.0}$   & 0.0 & $<5.7$ & 
                         0.1--14 \cr
$B^+\goto\etaprkstp$  & $11.1^{+12.7}_{-8.0}$ & 1.8 & $<35$ & 
                        1--3.7 \cr
$B^0\goto\etaprkstz$  & $7.8^{+7.7}_{-5.7}$   & 1.8 & $<24$ & 
                        1--8.0 \cr
$B^+\goto\etaprrhop$  & $11.2^{+11.9}_{-7.0}$ & 2.4 & $<33$ & 
                         3--24 \cr
$B^0\goto\etaprrhoz$  & $0.0^{+5.8}_{-0.0}$   & 0.0 & $<12$ & 
                         0.1--11 \cr
$B^+\goto\etak$       & $2.2^{+2.8}_{-2.2}$ & 0.8 & $<6.9$ &
                         0.2--5.0 \cr
$B^0\goto\etakz$      & $0.0^{+3.2}_{-0.0}$ & 0.0 & $<9.3$ &
                         0.1--3.0 \cr
$B^+\goto\etapi$      & $1.2^{+2.8}_{-1.2}$ & 0.6 & $<5.7$ &
                         1.9--7.4 \cr
$B^0\goto\etapiz$     & $0.0^{+0.8}_{-0.0}$ & 0.0 & $<2.9$ &
                         0.2--4.3 \cr
$B^+\goto\etakstp$    & \retaKstp & 4.8 &  see $\calB_{\rm fit}$    &
                         0.2--8.2 \cr
$B^0\goto\etakstz$    & \retaKstz & 5.1 &  see $\calB_{\rm fit}$    & 
                         0.1--8.9 \cr
$B^+\goto\etarhop$    & $4.3^{+5.2}_{-3.8}$ & 1.3 & $<15$ &
                         4--17 \cr
$B^0\goto\etarhoz$    & $2.6^{+3.2}_{-2.6}$ & 1.3 & $<10$ &
                         0.1--6.5 \cr
\end{tabular}
\end{center}
\label{combtab}
\end{table}

We have analyzed each of the decays without use of the likelihood
fit, employing more restrictive cuts in each of the variables to isolate
the signals.  The results are consistent with those quoted above
%in the tables, 
but with less precision.

\begin{figure}[htb]
\setlength{\epsfxsize}{1.0\linewidth}\leavevmode\epsfbox{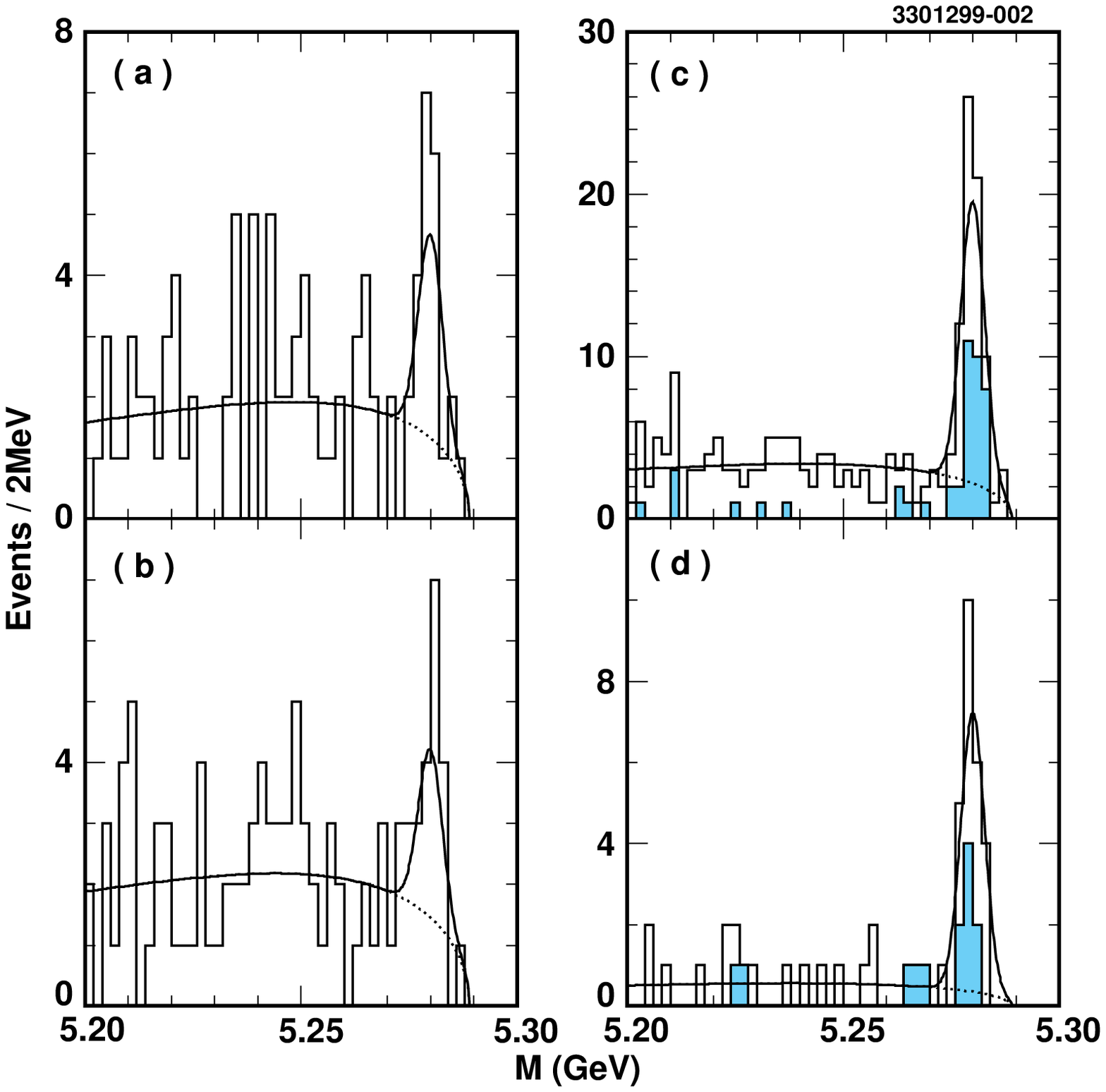}
\caption{\label{fig:mbproj}
Projections onto the variable \mb. The histograms show (a) $B^+\ra\eta K^{*+}$;
(b) \etaKstz; (c) $B^+\ra\eta^\prime K^+$; (d) \etapKz.  In (c) and (d)
the shaded histograms correspond to the $\eta^\prime\ra\eta\pi\pi$, 
$\eta\ra\gamma\gamma$ decay chain, while the unshaded histograms
correspond to the $\eta^\prime\ra \rho\gamma$ channel. 
The solid (dashed) line shows the 
projection for the full fit (background only) with the cut discussed in the 
text.}
\end{figure}

The signals we find in both charge states of \etaKst\ are first observations 
%\cite{chgconj}: $\BetaKstp\linebreak =\RetaKstp$ and $\BetaKstz=\linebreak
\cite{chgconj}: $\BetaKstp =\RetaKstp$ and $\BetaKstz=
\RetaKstz$.  
The significance, defined as the number of
standard deviations corresponding to the probability for a fluctuation
from zero to our observed yield,
is about 5 standard deviations for both.  We show
in Fig.\ \ref{fig:mbproj}\ the projections of event distributions
onto the \mb\ axis. A cut has been made to reject events with
small values of signal \calL, where for these purposes \calL\ is calculated
with \mb\ excluded.
The signals appear as peaks at the $B$ meson mass of 5.28 GeV in these plots. 
We also improve our previous measurements \cite{Betapr}\ of \etapK\ with 
the full CLEO II/II.V data sample:
$\BetapKp = \RetapKp$ and $\BetapKz= \RetapKz$.
The \mb\ projections for these modes are also shown in Fig.\ \ref{fig:mbproj}.

Assuming equal decay rates of charged and neutral $B$ mesons
to $\eta^{(\prime)} K^{(*)}$, we combine the measured
branching fractions \cite{chgneutfrac}. We obtain 
$\calB(B\ra\eta^\prime K) = (83^{+9}_{-8}\pm 7)\times10^{-6}$ and
$\calB(B\ra\eta K^{*}) = (18.0^{+4.9}_{-4.3} \pm 1.8)\times10^{-6}$.
We determine 90\% CL 
upper limits for $\calB(B\ra\eta^\prime K^*)$ and $\calB(B\ra\eta K)$
to be $22\times10^{-6}$ and $5.2\times10^{-6}$, respectively,
corresponding to central values of
$( 9.0^{+6.7}_{-5.0} )\times10^{-6}$ and 
$( 1.4^{+2.2}_{-1.4} )\times10^{-6}$ with 
statistical and systematic errors combined.
The pattern 
$\calB(\eta K) < \calB(\eta K^*) < \calB(\eta^\prime K)$
and $\calB(\eta^\prime K^*) < \calB(\eta^\prime K)$
is evident.

The observed branching fractions for \etapK\ and \etaKst, in combination
with the 
upper limits for the other modes in Table \ref{combtab} and with recent
measurements of $B\ra K\pi$, $\pi\pi$ \cite{kpipaper}, $B\ra\omega\pi$,
$\rho\pi$ \cite{PVLepPho99}, and $CP$ asymmetry in $B\ra K\pi$,
$\eta^\prime K$, $\omega\pi$ \cite{acpaper}
provide important constraints on the theoretical picture for these
charmless hadronic decays.  
The effective Hamiltonian calculations \cite{AKL,CCTY,theory} commonly used to
account for the charmless hadronic $B$ decays contain many uncertainties 
including form factors, light quark masses, CKM \cite{ckm} angles and the 
QCD scale.  
A large ratio of $\calB(\etapK,\eta K^*)$ to $\calB(\etaK,\eta^\prime K^*)$, 
consistent with our
measurements, was predicted qualitatively \cite{lipkin} in terms of
interference of the two penguin diagrams in Fig.\
\ref{fig:diagrams}(a) and (b), constructive for $\etapr K$ and $\eta K^*$
and destructive for $\eta K$ and $\etapr K^*$.  Most detailed calculations
\cite{AKL,CCTY,theory} predict a large branching fraction for the 
\etapK\ modes 
(though usually smaller than the observed values), but no enhancement
of \etaKst.  Three recent analyses \cite{GR,HSW,CY}, all of which take
guidance from charmless hadronic $B$ decay data, show that
the expectations for \etaKst\ can easily be  enhanced; the effective
Hamiltonian calculations accomplish this by increasing the relevant form factor
or decreasing the strange quark mass, the latter in
accordance with recent lattice calculations \cite{aoki}.  These and previous
calculations fall somewhat short of explaining the large rate for \etapK,
suggesting that the solution may involve contributions that are unique to
the \etapr\ meson.

We thank George Hou and Hai-Yang Cheng for many useful discussions.
We gratefully acknowledge the effort of the CESR staff in providing us with
excellent luminosity and running conditions.
This work was supported by 
the National Science Foundation,
the U.S. Department of Energy,
the Research Corporation,
the Natural Sciences and Engineering Research Council of Canada, 
the A.P. Sloan Foundation, 
the Swiss National Science Foundation, 
and the Alexander von Humboldt Stiftung.

\begin{thebibliography}{99}
\bibitem{chgneutfrac}
We assume the numbers of produced $B^+B^-$ and $B^0\bar B^0$ events to be 
equal.  The experimental uncertainty in the ratio is $\sim$10\%.
\bibitem{chgconj} 
Inclusion of the charge conjugate states is implied throughout this paper.
\end{thebibliography}
\end{document}